\documentclass[reprint,
 amsmath,amssymb,
 aps,
floatfix,
]{revtex4-2}

\usepackage{graphicx}
\usepackage{dcolumn}
\usepackage{bm}
\usepackage{mathtools}
\usepackage{lipsum}
\usepackage{physics}
\usepackage{braket}
\usepackage{subcaption}
\usepackage[justification=Justified, format=plain]{caption}
\usepackage{chemfig}
\usepackage{chemarrow}
\usepackage{tikz-cd}
\usepackage{comment}
\usepackage{hyperref}
\hypersetup{
    colorlinks=true,
    linkcolor=blue,
    filecolor=blue,      
    urlcolor=blue,
    pdftitle={Overleaf Example},
    pdfpagemode=FullScreen,
    citecolor=blue,
    }
\usepackage{orcidlink}

\begin{document}

\preprint{APS/123-QED}


\title{Intermediate time scale in the first product formation time  distribution of Michaelis-Menten kinetics with inhibitors}

\author{Arthur M. S. Carvalho \orcidlink{0009-0009-5279-1290}}
\email{arthurcarvalho1997@academico.ufs.br}
\author{Gerson C. Duarte-Filho \orcidlink{0000-0003-4975-4981}}%
\affiliation{%
 Departamento de Física, Universidade Federal de Sergipe, 49107-230 São Cristóvão, Sergipe, Brazil.}%

\author{Fernando A. N. Santos}
 \affiliation{Dutch Institute for Emergent Phenomena (DIEP),
Institute for Advanced Studies, University of Amsterdam, The Netherlands\\
Korteweg de Vries Institute for Mathematics, University of Amsterdam, The Netherlands }

\date{\today}

\begin{abstract}
Michaelis-Menten kinetics is one of the most recognized models in enzyme kinetics, crucial for the understanding of biochemical reactions in several metabolic processes. In this study, we perform a stochastic analysis of the Michaelis-Menten kinetics with the introduction of inhibitory mechanisms, which significantly diversifies the study of the reaction. We apply the Fock space formalism to reformulate the master equation, transforming it into a Schrödinger-type equation. 
We investigate reversible 
inhibitions and analyze the behavior of the averaged number of substances involved, identifying a stiffness behavior in all scenarios. In a specific case of partial inhibition, we observe that the inhibitor can act as an activator
that favors product formation. 
We calculate the first product formation time (FPFT), which characterizes the time statistic of the first product formation. 
We observe the emergence of an intermediate time scale in addition to the two known time scales typical in first-passage problems. This intermediate time scale is closely aligned with the slow-binding kinetics observed in experimentally observed enzymatic reactions that involve inhibitors. This intermediate time scale is related to the new pathways introduced by the presence of inhibitors.
This study 
offers a new perspective on inhibited enzymatic reactions, and demonstrates the usefulness of the Fock space formalism in the analysis of complex chemical systems.

\end{abstract}

\maketitle


\section{\label{sec:introduction}Introduction}




At the molecular level, where chemical kinetic reactions occur, the behavior of individual particles adheres to probabilistic laws \cite{Erdi2014Stochastic}. 
In this dynamic environment, chemical reactions are inherently stochastic, driven by chance encounters between reacting species.
The intrinsic stochastic nature poses a significant challenge in accurately characterizing the dynamics of such reactions, particularly in regimes where the number of substances is low \cite{Grima2014}.
In this regime, time measurements are noisy because of the random nature of the occurrence of these reaction events, thus requiring stochastics methods for their investigation.


The application of stochastic processes to chemical kinetics was pioneered in the 1940s. Kramers \cite{kramers1940brownian} conceptualized chemical reactions as the Brownian motion of particles, while Delbrück \cite{delbruck1940statistical} examined fluctuations in an autocatalytic model utilizing differential equations. 
In the late 1950s, Bartholomay \cite{bartholomay1958linear,*bartholomay1958stochastic,*bartholomay1959stochastic,*bartholomay1960molecular,*bartholomay1962enzymatic,*bartholomay1962stochastic} applied concepts of Markov chains to a birth-and-death process and chemical reactions, such as Michaelis-Menten chemical kinetics.
Over the years, similar methodologies were used to address more complex kinetic systems \cite{Erdi2014Stochastic}. 

In 1976, 
Masao Doi \cite{doi1976-1,*doi1976-2} demonstrated that the Fock space approach can be applied to classical stochastic processes. 
This implies that mathematical tools typically used in quantum mechanics can be applied to these processes. 
In this approach, the master equation 
is reformulated as a Schrödinger-type equation. 
The time evolution is governed by a quasi-Hamiltonian operator, 
written in terms of creation and annihilation ladder operators that describes all possible transitions among the system's configurations, along with a correction term to ensure normalization of probabilities \cite{baez_biamonte_2018}. 

This methodology 
provides an effective alternative to traditional stochastic simulation techniques, which require multiple simulations to obtain results, such as Gillespie's simulation methods \cite{gillespie:ARPC2007,Polizzi2016}. 
 Currently, such approach is being widely explored in various types of stochastic systems, such as absorbing states in non-equilibrium network dynamics \cite{dickman1991time}, spin chains \cite{alcaraz1994reaction}, gene expression \cite{sasai2003stochastic}, general reaction-diffusion dynamics \cite{isaacson2008relationship}, Michaelis-Menten chemical kinetics \cite{santos2015}, volume exclusion diffusion \cite{duarte2020}, Levy flights \cite{araujo2020revisiting,nicolau2021mean} and infectious disease modeling \cite{Mondaini2017,danilo2022}. 

In this work, we apply the Fock space approach to assess the role of the inhibition mechanism in the Michaelis-Menten kinetic reaction. 
In this reaction, the formation of the product is influenced by the presence of inhibitors that can adhere to the enzyme, the enzyme-substrate complex, or both. Here we consider the {\it competitive}, {\it uncompetitive}, {\it noncompetitive} and finally {\it partial} inhibition mechanisms \cite{klipp2016systems,bisswanger2017enzyme,cornish2013fundamentals,lehninger2014cox,thoma1960competitive}. 
The simple Michaelis-Menten model, although foundational, often fails to capture the complexities introduced by inhibitory mechanisms, leading to inaccurate parameter estimation and flawed predictions, particularly under physiological conditions or in the context of drug action \cite{leow_chang:BioPharm2019,fernandes_etal:AplSci2022,murugan:PO2024}. Inhibition is crucial for the regulation of metabolic pathways through feedback and is a key target for many drugs and toxins \cite{ramsay_etal:molecules2017}.

This study examines both algebraic and numerical methodologies for implementing the Fock space approach in the context of the Michaelis-Menten reaction that incorporates inhibition. 
Firstly, we consider a very simple situation in which only an enzyme, a substrate, and an inhibitor are considered. Simple analytical expressions are derived for the probabilities associated with all possible configurations.
With these probabilities at hand, we can calculate the moments associated with the number of a particular species at a given time. These mean values highlight the stiffness characteristic inherent in systems with multiple temporal scales \cite{gupta-khammash:EJP2014,kan-etal:JBM2016,BERTRAM2017105}. We further examined the scenario in which numerous particles from these species are capable of engaging in the reaction. 
In the case of many particles, specifically for partial inhibition, we observe that the inhibitor can act as a reaction activator, promoting product formation rather than hindering it. 



We treat the statistics associated with the time of the first formation of a product as a {\it first-passage} problem \cite{redner2001guide}. First-passage processes are essentially related to the probability that a particle in random motion reaches a specific location for the first time at a given time. The relevance of first-passage phenomena lies in their fundamental role in stochastic processes triggered by this initial event, covering several areas of study, such as chemical reactions \cite{szabo1980first,szabo1984localized,benichou2010geometry,scher2021unified,Polizzi2016}, enzymatic catalysis \cite{hopfield1974kinetic,ninio1987alternative,cao2011michaelis}, intracellular transport \cite{Bressloff-Newby:RMP2013}, molecular search processes \cite{eliazar2007searching,condamin2007first,benichou2011}, diffusion \cite{kou2003first,duarte2020}, to list just a few.
For more applications, see Ref. \cite{FPT_book:2014}. 

We obtain the first product formation time (FPFT) distribution for the Michaelis-Menten kinetic reaction with inhibition. Interestingly, an intermediate time scale is observed in the FPFT distribution for the competitive, noncompetitive and uncompetitive inhibitions. In general, the first-passage time (FPT) is described by two distinct time scales: one that delineates the dynamics at short time intervals and another that dictates the behavior in the long-term regime. 

Notably, the emergence of an intermediate decay regime in first-passage statistics is not unique to the inhibited Michaelis–Menten scheme.  Godec and Metzler showed that a diffusive searcher moving through a heterogeneous medium displays a three-stage FPT distribution, with a previously overlooked exponential segment that bridges the short-time power law and the asymptotic tail \cite{godec2016first}.  Single-molecule studies have reached similar conclusions: enzyme trajectories acquired in the presence of reversible inhibitors exhibit multiexponential waiting-time histograms that can be traced back to slow excursions between active and blocked conformations \cite{saha2012single}.  

More recently, Thorneywork et al. quantified first-passage events in colloidal escape, protein translocation through $\alpha$-hemolysin, and DNA-hairpin folding, and demonstrated that the short-time power-law exponent encodes the number of kinetic intermediates in each system \cite{thorneywork2020direct}.  At a fully theoretical level, 
it was proved that generic branched network topologies can give rise to hierarchies of time scales in first-passage problems \cite{li2013mechanisms,valleriani-etal:jcp2014}.

Placing our results in this broader context, we show that reversible inhibitor binding provides an elementary, analytically tractable route to the same phenomenology: the extra kinetic channel opens an intermediate time scale, whose magnitude follows directly from the subleading eigenvalues of the quasi-Hamiltonian.
We hypothesize that the emergence of this intermediate time scale is associated with intermediate pathways
caused by the interaction of the enzyme and the enzyme-substrate complex with inhibitors that delay the formation of the first product in our kinetic reaction.

This article is organized as follows. In Section \ref{sec:fock_space} we review the general formalism of the Fock space approach. The methodology is applied to Michaelis-Menten chemical kinetics with inhibitors in Section \ref{sec:michaelis_with_inhibitor}.
In Section \ref{sec:FPT}, we calculate the average number of particles of a specific substance participating in the reaction, determine the FPFT distribution and discuss the emergence of the intermediate time scale.
Finally, our main conclusions are summarized in Section \ref{sec:conclusion}.

\section{\label{sec:fock_space}The Fock space approach to mass action systems}

Suppose a typical reaction that follows the law of mass action. The reaction contains $L$ reactants $\mathcal{R}_i$, forming $M$ products $\mathcal{P}_j$ and has transition constants $k_{\pm}$, as described below:
\begin{equation}
     \sum_{i=1}^L \alpha_i \mathcal{R}_i(t) \overset{k_+}{\underset{k_-}\rightleftarrows} \sum_{j=1}^M \beta_j \mathcal{P}_j(t),
\end{equation}
where $\alpha_i$ and $\beta_j$ are stoichiometric coefficients. 
The probability of finding the system in a given configuration $\eta = \{\mathcal{R}_1,\dots, \mathcal{R}_L,\mathcal{P}_1,\ldots,\mathcal{P}_M\}$
in a time $t$ is represented by $P_{\eta}(t)$ 
and its time evolution is governed by the master equation
\begin{equation}
     \frac{\partial P_{\eta}(t)}{\partial t} = \sum_{\eta^{\prime}}[T_{\eta^{\prime} \to \eta}P_{\eta^{\prime}}(t) - T_{\eta \to \eta^{\prime}}P_{\eta}(t)],
     \label{eq:master_equation}
\end{equation}
where $T_{\eta^{\prime} \to \eta}$($T_{\eta \to \eta^{\prime}}$) denotes the transition rate between the states $\eta^{\prime}$($\eta$) and $\eta$($\eta^{\prime}$).

We define the Hilbert space $\boldsymbol{\mathcal{S}}_k = \{0,...,N_k\}$, where $k \in \{1,\ldots,L+M\}$
and $N_k$ is the maximum number of elements of the $k$th species.
A configuration state of the system in Fock space is represented as a direct product of all Hilbert spaces associated to the substances in the reaction, that is, 
$\boldsymbol{\mathcal{F}}= \boldsymbol{\mathcal{S}}_1 \bigotimes \ldots \bigotimes \boldsymbol{\mathcal{S }}_{L+M}$.
In Dirac notation, 
$\ket{\eta}=\ket{n_1\,n_2\,\ldots\, n_{L+M}}$
represents a pure Fock state, where $n_k \in \boldsymbol{\mathcal{S}}_k$ is the occupation number of each $k$th species.
The probability of finding the system in the state $\eta$ at time $t$
$P_\eta(t )$
is 
encoded in the state vector 
\begin{equation}
     \ket{\Psi(t)} := \sum_\eta P_\eta(t)\ket{\eta},
     \label{eq:fock_state}
\end{equation}
representing a linear superposition of the probability for each pure Fock state. 

We introduce the creation and annihilation ladder operators for each species acting on pure Fock states through the following equations:
\begin{subequations}
     \begin{equation}
         \boldsymbol{\alpha}^\dagger_j\ket{n} = \ket{n_1...(n_j+1)...n_k},
     \end{equation}
     \begin{equation}
         \boldsymbol{\alpha}_j\ket{n}=n_j\ket{n_1...(n_j-1)...n_k},
     \end{equation}
     \label{eq:operators}
\end{subequations}
\noindent that satisfy the commutation rules $[\boldsymbol{\alpha}_i,\boldsymbol{\alpha}_j^\dagger]=\delta_{ij}$,
and 
$[\boldsymbol{\alpha}_i,\boldsymbol{\alpha}_j]=[\boldsymbol{\alpha}_i^\dagger,\boldsymbol{\alpha}_j^\dagger]=0$ \cite{baez_biamonte_2018,mattis1998}
and $\boldsymbol{\alpha}_j^\dagger\boldsymbol{\alpha}_j$ corresponds to the number operator.

Inserting Eq. (\ref{eq:fock_state}) in the master equation Eq. (\ref{eq:master_equation}) 
we obtain the following Schrödinger-like equation 
\cite{doi1976-1,doi1976-2}:
\begin{equation}
     \frac{\partial \ket{\Psi(t)}}{\partial t} = -
     \boldsymbol{H} \ket{\Psi(t)},
     \label{eq:schrodinger_equation}
\end{equation}
where $\boldsymbol{H}$ is the quasi-Hamiltonian operator, written in terms of the ladder operators $\boldsymbol{\alpha}_j^\dagger$, $\boldsymbol{\alpha}_j$ and 
the rate constants. The solution of the 
Eq. (\ref{eq:schrodinger_equation}) 
is formally given by:
\begin{equation}
     \ket{\Psi(t)} = 
     \exp(-\boldsymbol{H}t)\ket{\Psi(0)},
     \label{eq:schrodinger_solution}
\end{equation}
where, $\ket{\Psi(0)}$ is the initial configuration. 

The quasi-Hamiltonian operator $\boldsymbol{H}$ is conveniently expressed as a square matrix whose dimension depends on the number of possible configurations of the system under study.
Using the Jordan normal form, Eq. (\ref{eq:schrodinger_equation}) can be written as:
\begin{equation}
     \ket{\Psi(t)} = \boldsymbol{Q}\exp(-\boldsymbol{J_H}t)\boldsymbol{Q^{-1}} \ket{\Psi(0)}.
     \label{eq:jordanform}
\end{equation}
with $\boldsymbol{J_H}$ being a diagonal matrix comprising the eigenvalues and $\boldsymbol{Q}$ being a matrix that contains the eigenvectors of $\boldsymbol{H}$.

In the next section, we use the Fock space approach to model the stochastic dynamics of Michaelis-Menten kinetics with inhibitors. We delineate the construction of the Fock space pertinent to this system and derive the precise solution for the scenario encompassing a single enzyme, substrate, and inhibitor, in addition to formulating the numerical procedures for cases involving multiple substances. 

\section{\label{sec:michaelis_with_inhibitor} Michaelis-Menten enzyme kinetics with inhibitors}

In traditional Michaelis-Menten kinetics, an enzyme $E$ binds to a substrate $S$ to form a complex $C_1$ ($ES$). This complex can either break down into the enzyme and substrate or be transformed into a product $P$ and an enzyme, in summary: 
\begin{equation}
    \begin{tikzcd}[column sep = large, row sep = huge, every label/.append style = {font = \fontsize{10}{10}\selectfont}]
        E + S \arrow[yshift=0.7ex]{r}{k_{1+}}
            & C_1 \arrow[yshift=-0.7ex]{l}{k_{1-}} \arrow[]{r}{k_2}
            & E + P,
    \end{tikzcd}
    \label{eq:michaelis}
\end{equation}
where $k_{1\pm}$ and $k_2$ are transition rates.
The Michaelis-Menten kinetics with inhibitor consists in introducing an inhibitor $I$ to the conventional reaction. The inhibitor can reversibly connect to the enzyme that forms the complex $C_2$ ($EI$) which, binding to the substrate, will form the new complex $ C_3$ ($EIS$). The other binding point of the inhibitor is with the $C_1$ complex which will also reversibly form the $C_3$ complex. Finally, the $C_3$ complex irreversibly transforms into product, enzyme, and inhibitor ($E+P+I$). The general scheme that describes the Michaelis-Menten kinetics with the inhibitor is:
\begin{equation}
    \begin{tikzcd}[column sep = huge, row sep = huge, every label/.append style = {font = \fontsize{10}{10}\selectfont}]
        E + S \arrow[yshift=0.7ex]{r}{k_{1+}} & C_1  \arrow[yshift=-0.7ex]{l}{k_{1-}} \arrow[]{r}{k_2} & E + P + I^*\\[-9ex]
        + \qquad & +\\[-9ex]
        I \qquad \arrow[xshift=-3ex, swap]{d}{k_{3+}} & I \arrow[xshift=-0.7ex, swap]{d}{k_{4+}}\\
        C_2 + S\arrow[yshift=0.7ex]{r}{k_{5+}} \arrow[xshift=-1.6ex, swap]{u}{k_{3-}} & C_3 \arrow[xshift=0.7ex, swap]{u}{k_{4-}} \arrow[yshift=-0.7ex]{l}{k_{5-}} \arrow[]{r}{k_6} & E + P + I
    \end{tikzcd}
    \label{eq:michaelis_inhibitor}
\end{equation}
The notation $I^*$ implies that $P$ is formed by the standard Michaelis-Menten mechanism. The $k_{1\pm}$, $k_2$, $k_{3\pm}$, $k_{4\pm}$, $k_{5\pm}$ and $k_{6}$ are rate constants. 

We shall define several specific types of inhibition that are considered in this study.
\begin{description}
\item[Competitive] The inhibitor can attach only to the enzyme and competes with the substrate for the single binding site on the enzyme. Under these conditions, $k_{4\pm}=k_{5\pm}=k_6=0$.
\item[Uncompetitive] The inhibitor binds only to the $C_1$ complex; the inhibitor does not compete with the substrate, since the enzyme has two distinct binding sites. In this situation, $k_{3\pm}=k_{5\pm}=k_6=0$.
\item[noncompetitive] The inhibitor can attach to both the enzyme $E$ and the complex $C_1$ without competition between them. Under this type of inhibition, $k_6=0$.
\item[Partial] Every reaction can occur as described in Eq. (\ref{eq:michaelis_inhibitor}), which means that the inhibitor only partially affects the reaction process. In this case, all rate constants are non-zero.
\end{description}

The Fock space for Michaelis-Menten kinetics with inhibitors is described as follows: 

\begin{equation}
\boldsymbol{\mathcal{F}} = \boldsymbol{\mathcal{S}} \bigotimes \boldsymbol{\mathcal{E}} \bigotimes \boldsymbol{\mathcal{I}} \bigotimes \boldsymbol{\mathcal{C}}_1 \bigotimes \boldsymbol{\mathcal{C}}_2 \bigotimes \boldsymbol{\mathcal{C}}_3 \bigotimes \boldsymbol{\mathcal{P}},
\end{equation}
The basis of this space is composed of kets in the form of 
$\ket{\eta} \equiv \ket{n_S\, n_E\, n_I\, n_{C_1}\, n_{C_2}\, n_{C_3}\, n_{P}}$,
where
$\eta = \{n_S, n_E, n_I, n_{C_1}, n_{C_2}, n_{C_3}, n_{P}\}$ 
represents a system configuration with
$n_S \in \boldsymbol{\mathcal{S}}$ substrates, $n_E \in \boldsymbol{\mathcal{E}}$ enzymes, $n_I \in \boldsymbol{\mathcal{I}}$ inhibitors, $n_{C_1} \in \boldsymbol{\mathcal{C}}_1$, $n_{C_2} \in \boldsymbol{\mathcal{C}}_2$ and $n_{C_3} \in \boldsymbol{\mathcal{C}}_3$ complexes $C_1$, $C_2$ and $C_3$, respectively, and $n_P \in \boldsymbol{\mathcal{P}}$ products.
These quantities must follow conservation laws for the number of substrates 
$N_S = n_S + n_{C_1} + n_{C_3} + n_P$, 
enzymes 
$N_E = n_E + n_{C_1} + n_{C_2} + n_{C_3}$
and inhibitors 
$N_I = n_I + n_{C_2} + n_{C_3}$
where, $N_S$, $N_E$, $N_I$ are the initial quantities of $S$, $E$, and $I$, respectively. 

The general state of the system in a given time $t$ is expressed by Eq. 
(\ref{eq:schrodinger_solution}),
with $\ket{\Psi(0)} = \ket{N_S, N_E, N_I, 0, 0, 0, 0}$ as the initial condition. 
The quasi-Hamiltonian $\boldsymbol{H}$ for this kinetics can be expressed using the creation and annihilation ladder operators as specified in Eq. (\ref{eq:operators})
and for the partial inhibition case reads

\begin{widetext}
\begin{eqnarray}
    \boldsymbol{H} &=& 
    k_{1+}(\boldsymbol{e}^{\dagger} \boldsymbol{s}^{\dagger}- \boldsymbol{c}^{\dagger}_1)\boldsymbol{es} + 
    k_{1-}( \boldsymbol{c}^{\dagger}_1 - \boldsymbol{e}^{\dagger} \boldsymbol{s}^{\dagger})\boldsymbol{c}_1 + 
    k_2(\boldsymbol{c}^{\dagger}_1-\boldsymbol{e}^{\dagger}\boldsymbol{p}^{\dagger})\boldsymbol{c}_1 + 
    k_{3+}( \boldsymbol{e}^\dagger   \boldsymbol{i}^\dagger - \boldsymbol{c}^\dagger_2)\boldsymbol{ei} + 
    k_{3-}(\boldsymbol{c}^\dagger_2 - \boldsymbol{e}^\dagger \boldsymbol{i}^\dagger)\boldsymbol{c}_2 \nonumber\\
    & & 
    + k_{4+}(  \boldsymbol{c}_1^\dagger \boldsymbol{i}^\dagger - \boldsymbol{c}_3^\dagger)\boldsymbol{c}_1\boldsymbol{i} + 
    k_{4-}( \boldsymbol{c}_3^\dagger - \boldsymbol{c}_1^\dagger \boldsymbol{i}^\dagger)\boldsymbol{c}_3 + 
    k_{5+}( \boldsymbol{c}_2^\dagger \boldsymbol{s}^\dagger - \boldsymbol{c}_3^\dagger)\boldsymbol{c}_2\boldsymbol{s} + 
    k_{5-}(  \boldsymbol{c}_3^\dagger - \boldsymbol{c}_2^\dagger \boldsymbol{s}^\dagger)\boldsymbol{c}_3 +
    k_6(  \boldsymbol{c}_3^\dagger - \boldsymbol{e}^\dagger \boldsymbol{p}^\dagger \boldsymbol{i}^\dagger)\boldsymbol{c}_3,
    \label{eq:quasi-hamiltonian}
\end{eqnarray}
\end{widetext}
where creation (annihilation) ladder operators 
$\boldsymbol{s}^\dagger$($\boldsymbol{s}$), $\boldsymbol{e}^\dagger$($\boldsymbol{e}$), $\boldsymbol{i}^\dagger$($\boldsymbol{i}$), $\boldsymbol{c}^\dagger_1$($\boldsymbol{c}_1$),  $\boldsymbol{c}^\dagger_2$($\boldsymbol{c}_2$), $\boldsymbol{c}^\dagger_3$($\boldsymbol{c}_3$) and  $\boldsymbol{p}^\dagger$($\boldsymbol{p}$), correspond to the substrate, enzyme, inhibitor, $C_1$, $C_2$ and $C_3$ complexes and product, respectively. 
This expression accounts for different types of inhibition, including the case without any inhibitors. Specifically, when $k_{3\pm}=k_{4\pm}=k_{5\pm}=k_6=0$, it models the Michaelis-Menten kinetics without inhibitors as discussed in Ref. \cite{santos2015}. In the case where $k_{4\pm}=k_{5\pm}=k_6=0$, it corresponds to competitive inhibition. When $k_{3\pm}=k_{5\pm}=k_6=0$, it deals with uncompetitive inhibition. Lastly, with $k_6=0$, 
represents noncompetitive inhibition.

\subsection{\label{sub:single_michaelis}Single enzyme-substrate-inhibition analytical solutions}


We illustrate a straightforward example to demonstrate the adaptability of the Fock space methodology to study chemical reactions. We consider the uncompetitive inhibition for this purpose, characterized by $k_{3\pm}=k_{5\pm}=k_6=0$ in Eq. (\ref{eq:michaelis_inhibitor}). In this specific inhibition, there is no formation of the complex $C_2$, therefore, we use the following reduced Fock space \(\boldsymbol{\mathcal{F}}_{\rm uncomp} = \boldsymbol{\mathcal{S}} \bigotimes \boldsymbol{\mathcal{E}} \bigotimes \boldsymbol{\mathcal{I}} \bigotimes \boldsymbol{\mathcal{C}}_1 \bigotimes  \boldsymbol{\mathcal{C}}_3 \bigotimes \boldsymbol{\mathcal{P}}\) spanned by a complete basis formed by the kets \(\ket{\eta}=\ket{n_S\,n_E\,n_I\,n_{C_1}\,n_{C_3}\,n_P}\). 
For $N_S=N_E=N_I=1$, there are four possible states

\begin{eqnarray*}
	& &\ket{1}=\ket{111000}, \ket{2}=\ket{001100}, \ket{3}=\ket{000010},\\
	& &\ket{4}=\ket{011001}, 
\end{eqnarray*}
where the chosen initial condition is $\ket{\Psi(0)}=\ket{1}$. 
It should be noted that since the reaction $E+S \to P$ is the sole irreversible reaction, the system dynamics ceases when the state $\ket{4}$ is reached. In the context of first-passage processes, this state is referred to as an {\it absorbing state}. This reaction will subsequently be addressed as a first-passage problem. 

To find the matrix representation of the quasi-Hamiltonian, we determine the elements of the matrix 
$\bra{\eta^{\prime}}\boldsymbol{H}\ket{\eta} \equiv H_{\eta^{\prime}\eta}$
using the states described above. Since these states are orthonormal, we have 
$\braket{\eta^\prime|\eta}=\delta_{\eta^\prime,\eta}$. Hence, the matrix representation of 
the quasi-Hamiltonian is given by:

\begin{equation}
    \boldsymbol{H} = \left(
    \begin{matrix}
        k_{1+} & -k_{1-} & 0  & 0\\
        -k_{1+} & k_{1-}+k_2+k_{4+} & -k_{4-} & 0  \\
        0 & -k_{4+} & k_{4-} & 0  \\
        0 & -k_{2} & 0 & 0 \\
    \end{matrix}
    \right).
    \label{eq:matrix_H}
\end{equation}

To make the analysis easier, we set
$k_{1 \pm} = k_{4 \pm} \equiv k^{\prime}$ and $k_2 \equiv k$. This shortens the analytical expressions, yet retains compelling characteristics of this stochastic reaction.
The characteristic polynomial is defined as

\begin{equation}
    p(\lambda) = \lambda  (\lambda - k^{\prime})
    \left[\lambda^2 - (k+3k^{\prime})\lambda + kk^{\prime}\right] ,
\end{equation}
where the associated eigenvalues are:
$\lambda_1=0$, $\lambda_2=k^{\prime}$ and $\lambda_{3,4}= (k+3k^{\prime }\pm \sqrt{\Delta})/2$, where $\Delta = k^2 + 2kk^{\prime} + 9k^{\prime 2}$.
Hence, the diagonal matrix consisting of the exponential of the eigenvalues of $\boldsymbol{H}$ is given by
$\exp(-\boldsymbol{J_H}t)=
\boldsymbol{{\rm diag}} \left(1,e^{-\lambda_2t},e^{-\lambda_3t},e^{-\lambda_4t}\right)$
Each column of the matrix $\boldsymbol{Q}$ of Eq. (\ref{eq:jordanform}) represents a normalized eigenvector of $\boldsymbol{H}$. Although the matrix $\boldsymbol{H}$ is not highly dimensional, the length of the resulting eigenvectors is significant, making it impractical to explicitly present the matrix $\boldsymbol{Q}$ here.
Using Eq. (\ref{eq:jordanform}) we obtain a linear combination of the Fock states
\begin{equation}            
    \ket{\Psi(t)} = P_1(t)\ket{1} + P_2(t)\ket{2} + P_3(t)\ket{3} + P_4(t)\ket{4},
\end{equation}
where 
$P_1(t)$, $P_2(t)$, $P_3(t)$, $P_4(t)$, are the probabilities of the occurrence of states $\ket{1}$, $\ket{2}$, $\ket{3}$,
and $\ket{4}$, respectively, at time $t$. The probabilities are given by:


\begin{widetext}
    \begin{eqnarray}
        P_1(t) &=& \frac{1}{2}e^{-\frac{1}{2}(k+3k')t}\left[ \frac{(k+k')}{\sqrt{\Delta}}\sinh\left(\frac{\sqrt{\Delta}t}{2}\right) + \cosh\left(\frac{\sqrt{\Delta}t}{2}\right) + e^{\frac{1}{2}(k+k')t} \right]  ,
    \\
        P_2(t) &=& \frac{2 k' e^{-\frac{1}{2} (k+3k')t}}{ \sqrt{\Delta }} \sinh \left(\frac{\sqrt{\Delta } t}{2}\right), 
    \\
        P_3(t) &=& \frac{1}{2}e^{-\frac{1}{2}(k+3k')t}\left[ \frac{(k+k')}{\sqrt{\Delta}}\sinh\left(\frac{\sqrt{\Delta}t}{2}\right) + \cosh\left(\frac{\sqrt{\Delta}t}{2}\right) - e^{\frac{1}{2}(k+k')t} \right]  ,
    \\
        P_4(t) &=& 1 - \left[ \frac{(k+3k')}{\sqrt{\Delta}}\sinh\left(\frac{\sqrt{\Delta}t}{2}\right) + \cosh\left(\frac{\sqrt{\Delta}t}{2}\right) \right]e^{-\frac{1}{2}(k+3k')t}.
    \end{eqnarray}
    \label{eq:probabilities}
\end{widetext}

The $l$th moment of the amount of substance $\sigma$ in a time $t$ is given by
\begin{equation}
    \langle n_\sigma^l(t) \rangle = \sum_{\eta}n^l_\sigma P_\eta(t), \hspace{0.5cm} l\in \mathbb{N},
    \label{eq:mean}
\end{equation}
where, $n_\sigma$ is the amount of such substance
in configuration $\eta$ and $\sigma=\{S,E,I,C_1,C_2,C_3,P\}$. The first and second moments, denoted by $\langle n_\sigma(t) \rangle$ and $\langle n_\sigma^2(t) \rangle$, respectively, are equal when $N_S=N_E=N_I=1$. For the uncompetitive inhibition, they are given by: 
$\langle n_S(t) \rangle=P_1(t)$, $\langle n_E(t)\rangle= P_1(t)+P_4(t)$,  $\langle n_I(t) \rangle=P_1(t)+P_2(t)+P_4(t)$, $\langle n_{C_1}(t)\rangle=P_2(t)$, $\langle n_{C_3}(t)\rangle=P_3(t)$ and $\langle n_P(t)\rangle=P_4(t)$.

\begin{figure}[h]
    \includegraphics[width=\columnwidth]{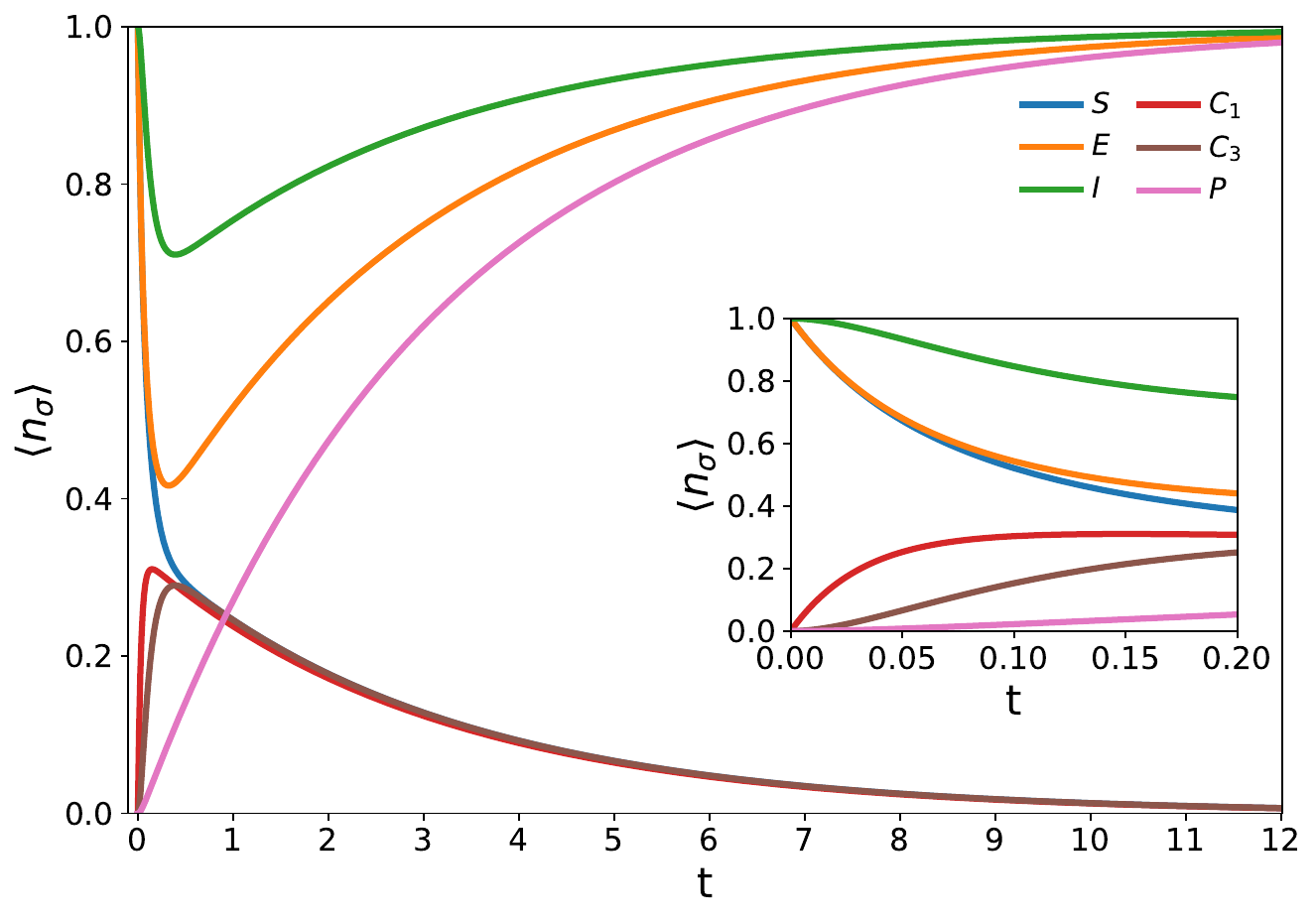}
    \caption{\label{fig:mean_single} Temporal evolution of mean quantities $\langle n_\sigma(t)\rangle$ for uncompetitive inhibition, with $N_S=N_E=N_I=1$ and rate constants $k'=10$ and $k=1$. Inset: magnified section of the main plot for short time depicting the stiffness inherent in this system.}
\end{figure}

Figure \ref{fig:mean_single} shows the time evolution of the mean quantities $\langle n_\sigma (t) \rangle$ for 
uncompetitive inhibition. 
It is evident that these quantities approach their expected values for $t \gg 1$, namely \(\langle n_P(t) \rangle = \langle n_E(t)  \rangle = \langle n_I(t) \rangle \to 1\) and \(\langle n_S(t) \rangle = \langle n_{C_1}(t)  \rangle  = \langle n_{C_3}(t) \rangle \to 0\). 
Figure \ref{fig:mean_single} also shows in its inset rapid changes in these averaged quantities on short timescales $t\approx0$, unlike the stable behavior observed on long timescales. This behavior is known as stiffness \cite{murray2002mathematical,santos2015}. 
Stiffness usually causes inefficiencies in stochastic simulations.
As one can see, the Fock space approach efficiently handles stiff systems, rendering this method a viable alternative for stochastic simulations.

We address the inherent stiffness of this system by examining the mean value of $S$.
We can identify two distinct time scales for $\langle n_S(t) \rangle$. Expanding $\langle n_S(t) \rangle$ 
around $t=0$ and neglecting terms of second order or higher, 
we get 
$\langle n_S(t) \rangle \approx (1 - k't) \approx e^{-k't}$, so we have $\langle n_S(t \approx 0) \rangle \approx e^{-t /\tau_1}$, where 
$\tau_1= 1/k'=1/\lambda_2$ is the first time scale. For $t \gg 1$, we have $\cosh{(\frac{\sqrt{\Delta}}{2}t)}=\sinh{(\frac{\sqrt{\Delta}}{2}t)}\approx e^{\frac{\sqrt{\Delta}}{2}t}/2$, 
that gives $\langle n_S(t \gg 1) \rangle \propto e^{-t/\tau_2}$, where 
$\tau_2=1/\lambda_4$.
In summary, the average number of $S$ decreases more quickly over short timescales, dictated by $\tau_1$, while over longer timescales the decrease is slower and is defined by $\tau_2$. This highlights another inherent characteristic of such a system: the presence of multiple timescales. 
Computer simulations of multi-timescale stochastic reactions, using, for example, the Gillespie algorithm \cite{gillespie:ARPC2007}, can be very demanding in resolving fastest timescales \cite{gupta-khammash:EJP2014,kan-etal:JBM2016,BERTRAM2017105}. 
Our method efficiently handles these systems and requires minimal computational resources.


\subsection{\label{sub:multiple_michaelis} Solutions for a finite number of enzymes, substrates, and inhibitors}

We extend our analysis to a more complex scenario that involves multiple enzymes, substrates, and inhibitors interacting simultaneously. 
The system comprises a total of 12 distinct substances: 6 substrates ($N_S=6$), 3 enzymes ($N_E=3$), and 3 inhibitors ($N_I=3$).
Due to the increase in the number of substances, the number of possible states grows very rapidly depending on the type of inhibition considered. 
For partial inhibition, 
there are 110 possible states, 
making the analysis of $\exp(-\boldsymbol{H}t)$ using the Jordan form unfeasible. 

In such situations, we resort to a numerical solution for $\exp(-\boldsymbol{H}t)$ using the \texttt{expm()} function of the \texttt{Scipy} library of the \texttt{Python} programming language. It is a robust tool for the numerical computation of matrix exponentials using Higham's algorithm \cite{Higham2005}. In certain instances, the quasi-Hamiltonian $\boldsymbol{H}$ assumes a sparse matrix structure. Using this sparsity can significantly reduce computational overhead when evaluating $\exp(-\boldsymbol{H}t)$ through appropriate numerical methodologies tailored for sparse matrices.

\begin{figure}[h]
    \includegraphics[width=\columnwidth]{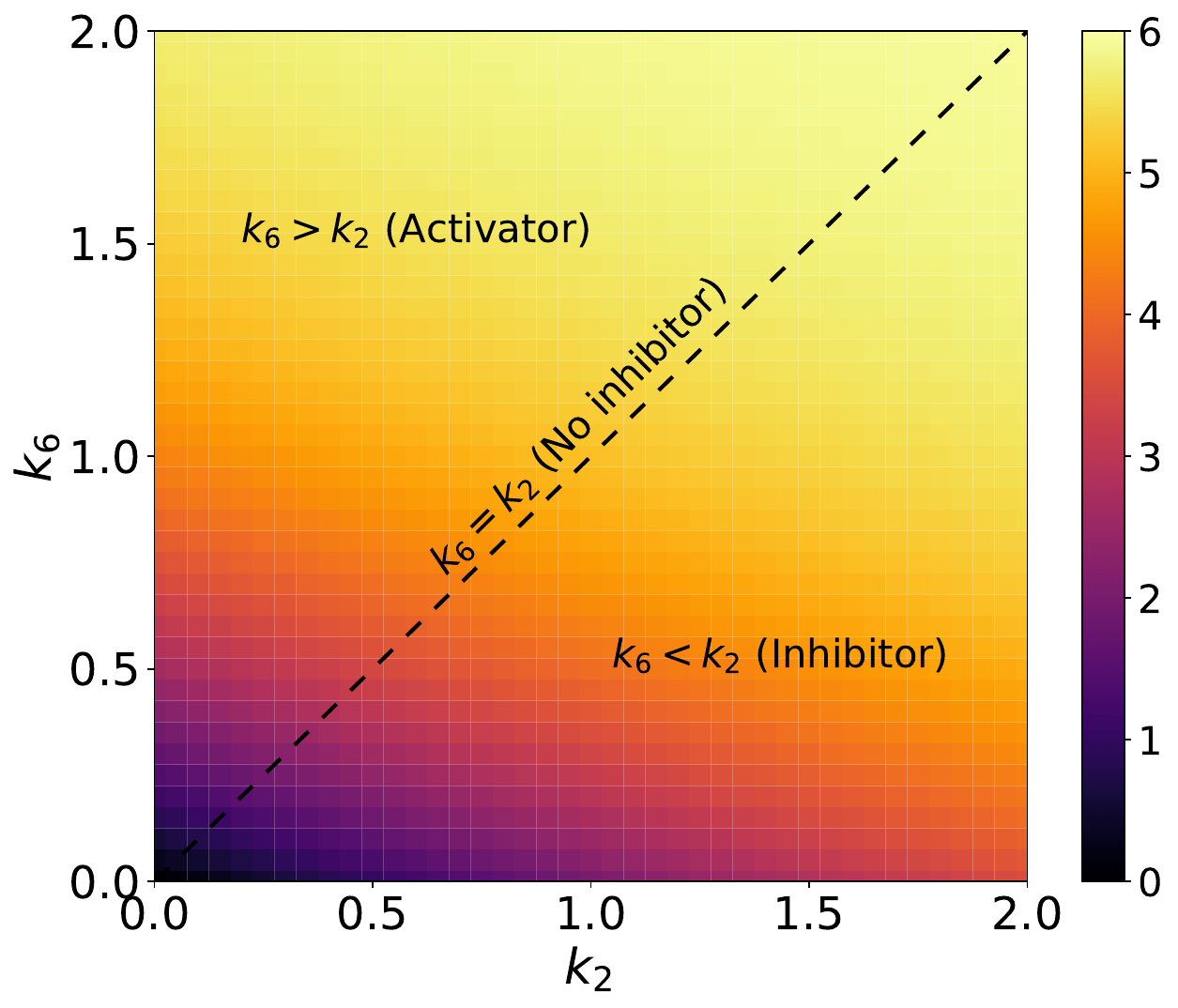}
    \caption{\label{fig:product_partial} Mean number of formed product in a partial inhibition as a function of $k_2$ and $k_6$ for $N_S=6$, $N_E=N_I=3$,   $k_{1+}=k_{3+}=k_{5+}=k_{4+}=10$ and $k_{1-}=k_{3-}=k_{4-}=k_{5-}=5$ at $t=3$. The dashed line $k_2=k_6$ separates the space parameters in two regions: $k_2>k_6$, where the inhibitor hinders the formation of the product; $k_2<k_6$, where the inhibitor functions as an activator.}
\end{figure}

To implement our numerical routine, we represent 
the elements 
$H_{\eta^{\prime}\eta}$ using Kronecker deltas
as follows


\begin{equation} 
\begin{split}
    H_{\eta^{\prime}\eta} & = \sum_{\tau} \tau \left(  \prod_\sigma f(\Delta_{\sigma,\tau})\delta_{n_\sigma^\prime,n_\sigma+\Delta_{\sigma,\tau}} - \right. \\
    &  \left. \prod_\sigma f(\Delta_{\sigma,\tau})\delta_{n_\sigma ',n_\sigma} \right),
\end{split}
    \label{eq:hamiltonian-kronecker}
\end{equation}
where $\tau=\{k_{1+}, k_{1-},\dots\}$ is the set of reactions.
The factor $\Delta_{\sigma,\tau}=0,\pm 1$ describes the transitions. 
For example, consider the reaction $k_{1+}$ when $S$ and $E$ form $C_1$, we have $\Delta_{\{S,E\},k_{1+}}=-1$ for 
$S$ and $E$, $\Delta_{C_1,k_{1+}}=1$ for 
$C_1$ and $\Delta_{\bar{\sigma},k_{1+}}=0$ for the other substances $\bar{\sigma} = \{I,C_2,C_3,P\}$. 
The term $f(\Delta_{\sigma,\tau})$ is an auxiliary function that provides the exact multiplicity of substance $\sigma$ in reaction $\tau$ and reads

\begin{equation}
    f(\Delta_{\sigma,\tau}) =
    \begin{cases}
        1 & \text{if } \Delta_{\sigma,\tau} = 0 \text{ or } 1, \\
        n_{\sigma} & \text{if } \Delta_{\sigma,\tau} = -1.
    \end{cases}
\end{equation}
Figure \ref{fig:product_partial} shows a heatmap of the average amount of product in partial inhibition as a function of $k_2$ and $k_6$ for $t=3$. 
We consider $N_S=6$ and $N_E=N_I=3$, with rate constants $k_{1+}=k_{3+}=k_{5+}=k_{4+}=10$ and $k_{1-}=k_{3-}=k_{4-}=k_{5-}=5$.
The black dashed line represents the situation $k_6=k_2$, where the formation of the product has the same kinetic behavior without inhibitor. 
In the region above the dashed line, $k_2<k_6$, the inhibitor functions as an activator, whereas below it, $k_2>k_6$, the inhibitor hinders the formation of $P$. $\langle n_P(t) \rangle$ is greater in region $k_2<k_6$ than in region $k_2>k_6$ for a fixed value of $k_2$. Fig. \ref{fig:product_partial} also stresses the efficiency of the Fock space approach compared with numerical simulations, since it enables us to construct parameter space without requiring demanding ensemble averaging, especially for multi-timescale stiff systems. 

To build the heatmap in Fig. \ref{fig:product_partial} we performed a total of 1600 calculations of $e^{-\boldsymbol{H}t}$ by varying $k_2$ and $k_6$ from 0 to 2, with an increment of $\text{inc}=0.05$ per step and took a time of the order of ten minutes to obtain on a personal computer. 
This is due to the remarkable efficiency of the Scipy routine used in our computations, which benefits significantly from the sparse structure of the matrix \(\boldsymbol{H}\), particularly when a limited set of substances is considered.
This offers a notable benefit over computational simulations, which require considerably more time.
Using the CAIN software \cite{stalzer-mauch:CAIN2011}, which employs Gillespie's direct method, to execute identical tasks on the same personal computer would require a duration approximately 20 times longer. 

It should be noted that, with an increase in the number of substances, the efficiency of the Fock space approach diminishes, ultimately rendering stochastic simulations more advantageous. 
In our analysis, this trend is observed when the total number of substances involved in the reactions exceeds 16 substances.  
As the matrix becomes increasingly dense, the efficiency observed initially is no longer maintained. 
Naturally, strategies could be developed to address the challenges posed by nonsparse matrices, with the aim of enhancing the performance of the Fock space approach when dealing with reactions involving a greater number of substances. 
Nevertheless, it is crucial to emphasize that the methodology delineated herein is intended for the examination of small stochastic systems. Such systems, despite their limited scale, often exhibit significant complexity due to extensive parameter spaces, which in turn pose challenges in inference, parameter estimation, and model identifiability \cite{gupta-khammash:SIAM2013}.

\section{\label{sec:FPT} First Product Formation Time Distribution}




In this section, we discuss the problem of determining the time required for the formation of the first product in the Michaelis-Menten kinetics with inhibition as a first-passage process \cite{redner2001guide}.

\subsection{First Product Formation as a First-Passage Process}

Consider once again the simplest case $N_E=N_S=N_I=1$. 
Since the reaction $E+S \to P$ is irreversible, once the system reaches the state $\ket{4} = \ket{011001}$, the product is formed, and the reaction is over. 
We define $T$ as the random variable for the first product formation time (FPFT).
The cumulative probability 
that the first product formation occurs for $T \leq t$ is given by 
$$\mathcal{P}(T\leq t) = 1 - \sum_{\eta \neq \{\eta_{\rm FP}\}} P_{\eta}(t):= \mathcal{F}(t),$$
where $\{\eta_{\rm FP}\}$ is the set of all configurations with at least one product.
The probability distribution of the FPFT is the derivative of $\mathcal{F}(t)$
\begin{equation}
    f(t) = \frac{\partial \mathcal{F}(t)}{\partial t} = - \sum_{\eta \neq \{\eta_{\rm FP}\}} \frac{dP_{\eta}(t)}{d t}.
 \label{eq:FPT_fock}
\end{equation}
Substituting the master equation, Eq. (\ref{eq:master_equation}) in Eq. (\ref{eq:FPT_fock}) one gets
\begin{equation}
    f(t) = \sum_{\eta \neq \{\eta_{\rm FP}\}} \sum_{\eta'} \left[ T_{\eta \to \eta '}P_\eta (t)-T_{\eta' \to \eta} P_{\eta'}(t)\right].
\end{equation}
The only terms that survive from the above summations are those related to transitions from configurations $\eta \neq \{\eta_{\rm FP}\}$ to $\eta^{\prime} = \{\eta_{\rm FP}\}$, thus
\begin{equation}
    f(t) = \sum_{\eta \neq \{\eta_{\rm FP}\}} \sum_{\eta' = \{\eta_{\rm FP}\}} T_{\eta \to \eta '}P_\eta (t). 
    \label{eq:FPT_general}
\end{equation}

\begin{figure}
    \centering
    \includegraphics[width=0.48\textwidth]{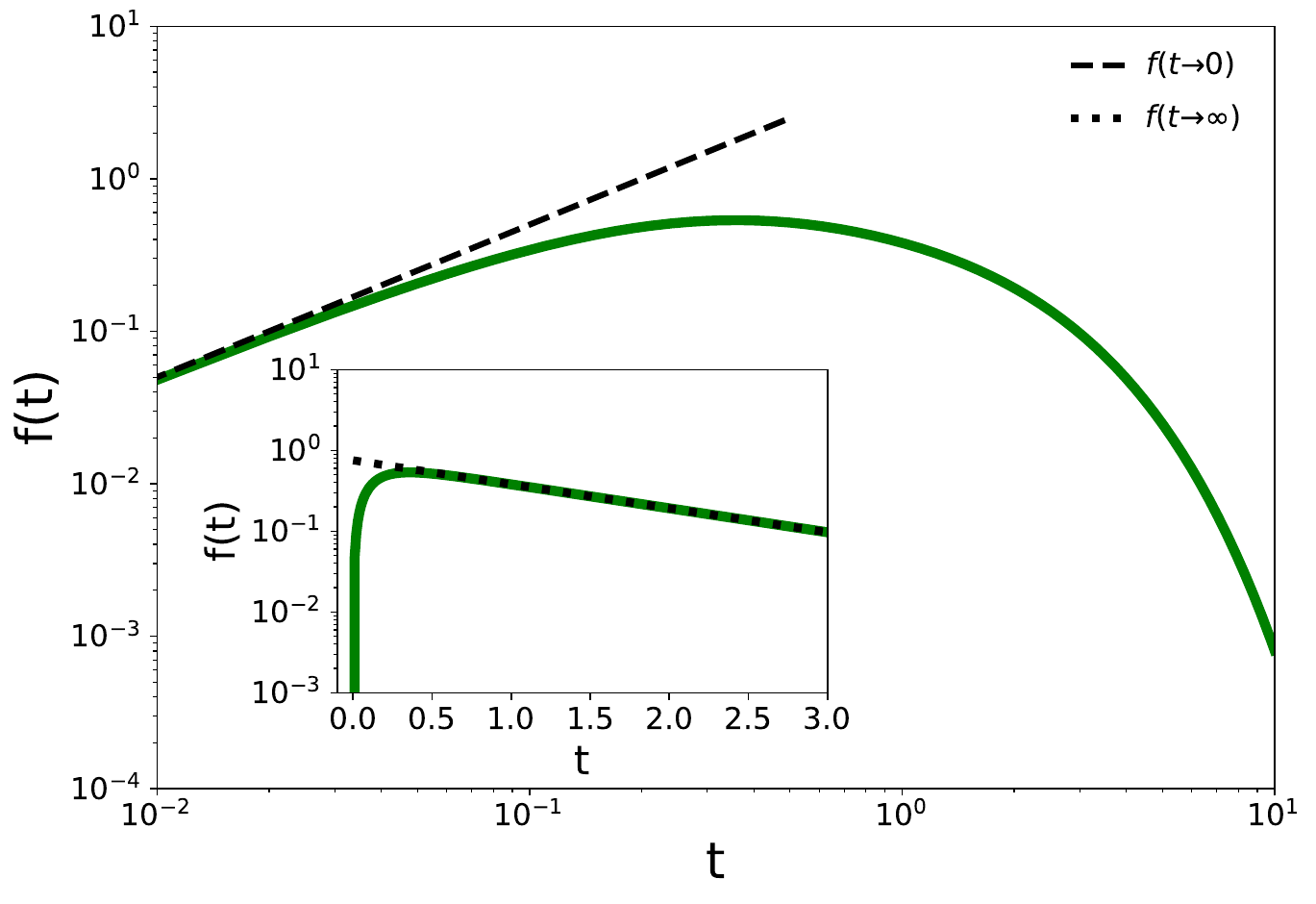}
    \caption{
    The probability distribution of the first product formation time for uncompetitive inhibition, Eq. (\ref{eq:FPT_uncomp_simplest}), with  $N_S = N_E = N_I = 1$, $k = 5$ and  $k' = 1$ on a log-log scale.  The dashed line represents the linear behavior for  $t \to 0$, while the dotted line shown in the inset on a log-normal scale represents the behavior for  $t\gg 1$.}
    \label{fig:FPT_partial_single}
\end{figure}

For the simplest case treated in Subsection \ref{sub:single_michaelis}, Eq. (\ref{eq:FPT_general}) reads

\begin{equation}
    f(t) = T_{2\to4} P_2(t) = \frac{2 kk' e^{-\frac{1}{2} (k+3k')t}}{ \sqrt{\Delta }} \sinh \left(\frac{\sqrt{\Delta } t}{2}\right).
    \label{eq:FPT_uncomp_simplest}
\end{equation}
Note that this is exactly the average amount of $C_1$ multiplied by the rate of irreversible transition $k$. For short times $t \to 0$, $f(t) \approx kk't$, while for $t\gg1$ $f(t) \approx (kk'/\sqrt{\Delta})e^{-\lambda_4t}$.
Figure \ref{fig:FPT_partial_single} illustrates the behavior of the FPFT distribution for 
uncompetitive inhibition Eq. (\ref{eq:FPT_uncomp_simplest})
with 
$k=5$ and $k_+ = k_- = 1$. The dashed line represents the linear behavior of $f(t)$ for $t \to 0$ on a logarithmic scale. The dotted line represents the linear behavior of the tail of the distribution in the inset on a logarithmic normal scale.

For the general case, where $N_S, N_E, N_I > 1$, the distribution $f(t)$ is given by
\begin{equation}
    f(t)=k_2\braket{n_{C_1}(t)}_{\eta\neq\{\eta_{\rm PF}\}} + k_6\braket{n_{C_3}(t)}_{\eta\neq\{\eta_{\rm PF}\}},
    \label{eq:FPT_multiple}
\end{equation}
where the average quantities of $C_1$ and $C_3$ are performed considering all configurations except those where at least one product was formed. 
In the following, we use Eq. (\ref{eq:FPT_multiple}) to obtain the mean formation time of the first product and its relationship with the reaction rate.



\subsection{Mean First Product Formation Time}

The $l$th moment of the FPFT distribution is obtained as follows 
\begin{equation}
     \langle T^l \rangle = \int_0^\infty t^l f(t)dt.
     \label{eq:MFPT}
\end{equation}
For $l=1$, we have the mean first product formation time $\braket{T}$.
From a chemical kinetic perspective, the reaction rate $\nu$ is proportional to the inverse of the mean first product formation time $\braket{T}$ \cite{kumar2015,Polizzi2016}.
The reaction rate indicates how quickly a product is formed. 

\begin{figure}
    \includegraphics[width=\columnwidth]{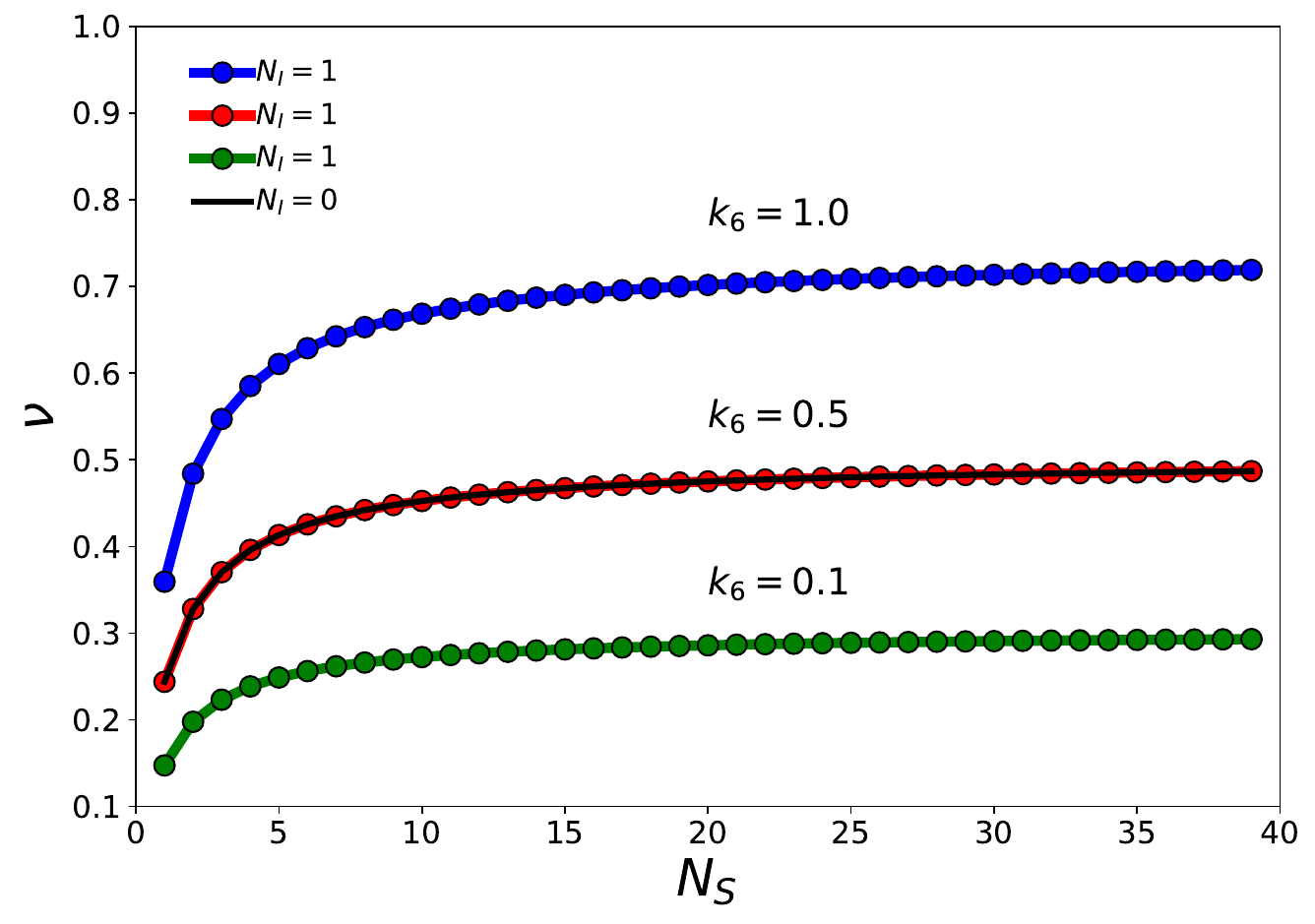}
    \caption{\label{fig:reaction_rate} Reaction rate for partial inhibition with varying values of $k_6$, $N_E = 1$ and $k_{1\pm} = k_{3\pm} = k_{4\pm} = k_{5\pm} = 10$ and $k_2 = 0.5$. The colored lines represent the reaction rate for partial inhibition, whereas the black line corresponds to the Michaelis-Menten reaction without inhibitors.}
\end{figure}

Figure \ref{fig:reaction_rate}  presents the reaction rate as a function of 
$N_S$, with $N_E=N_I=1$, 
$k_{1\pm}=k_{3\pm}=k_{4\pm}=k_{5\pm}=10$, $k_2=0.5$ and $k_6=0.1$ (green curve), $0.5$ (red) and $1.0$ (blue) for the partial inhibition. The usual Michaelis-Menten kinetic reaction, $N_I=0$, is presented as a solid black curve.   
For $k_2 = k_6$, the partial inhibition is equivalent to the usual Michaelis-Menten reaction, a behavior also illustrated in Figure (\ref{fig:product_partial}). 
Furthermore, when $k_2 < k_6$, the inhibitor increases the reaction rate, while for $k_2 > k_6$, the reaction rate is lower, resulting in 
delay in product formation.

\begin{figure}
    \includegraphics[width=\columnwidth]{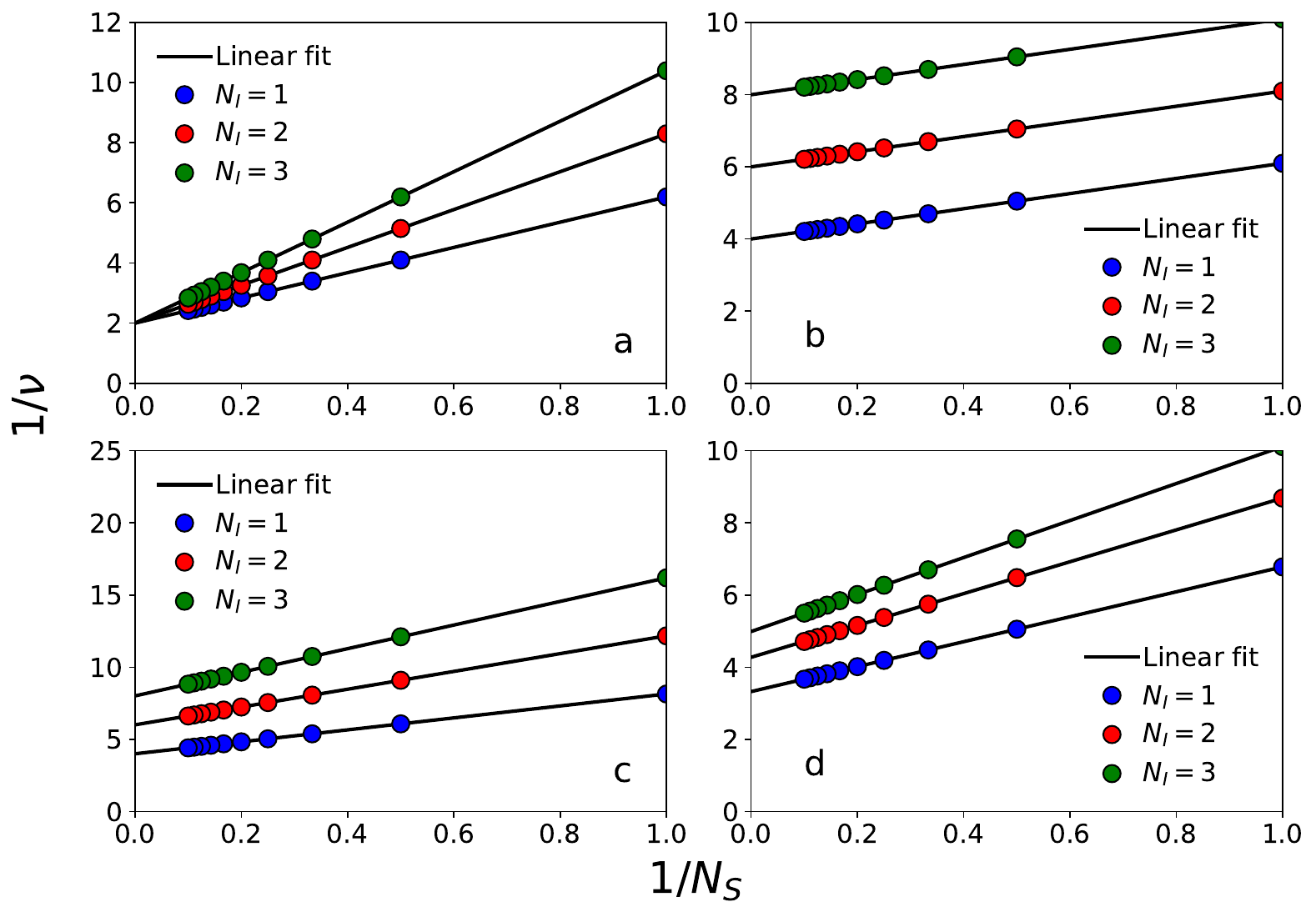}
    \caption{\label{fig:LB_inhibition} Lineweaver-Burk linearization for the different inhibitions with $N_E=1$, 
    $k_{1\pm}=k_{3\pm}=k_{4\pm}=k_{5\pm}=10$, $k_2=0.5$ and $k_6=0.1$. (a) competitive inhibition, (b) uncompetitive inhibition, (c) noncompetitive inhibition,  and (d) partial inhibition. The points were obtained using Eq. (\ref{eq:FPT_multiple}) and the black line is a linear fit of the points.}
\end{figure}

Lineweaver-Burk linearization (LBL) is performed by plotting $1/\nu$ versus $1/N_S$ \cite{lineweaver-burk1934}. 
The slope and $y$-intercept of this linearization, which can be determined experimentally, yield important kinetic parameters of the reaction.
Figure \ref{fig:LB_inhibition} shows the LBL for different types of inhibitions, with $N_E=1$, 
$k_{1\pm}=k_{3\pm}=k_{4\pm}=k_{5\pm}=10$, $k_2=0.5$ and $k_6=0.1$.
For competitive inhibition, shown in panel (a), the slope of the line varies as the amount of inhibitor increases. In the case of uncompetitive inhibition, panel (b), only the linear coefficient changes. However, for noncompetitive and partial inhibitions, panels (c) and (d), respectively, both linear and angular coefficients vary.

These results are in agreement with previous studies that analyzed the Michaelis-Menten reaction with inhibitors \cite{klipp2016systems,bisswanger2017enzyme,cornish2013fundamentals,lehninger2014cox,thoma1960competitive}. 
The fact that a LBL only requires two data points to establish a linear relationship offers a powerful validation strategy for our stochastic method, which is particularly efficient with a small number of reactants. We can quickly construct the linearization using just two data points generated by our method with $N_S=1, 2$. The resulting slope can then be efficiently compared with results obtained from numerical methods designed for the large copy number regime, providing a quick check on the precision of our approach in the small copy number limit.

\subsection{Intermediate Time Scale}

\begin{figure*}
  \begin{tabular}{cc}
    \subcaptionbox{Competitive\label{fig:FPT-competitive}}[0.47\linewidth]{\includegraphics[width=\linewidth]{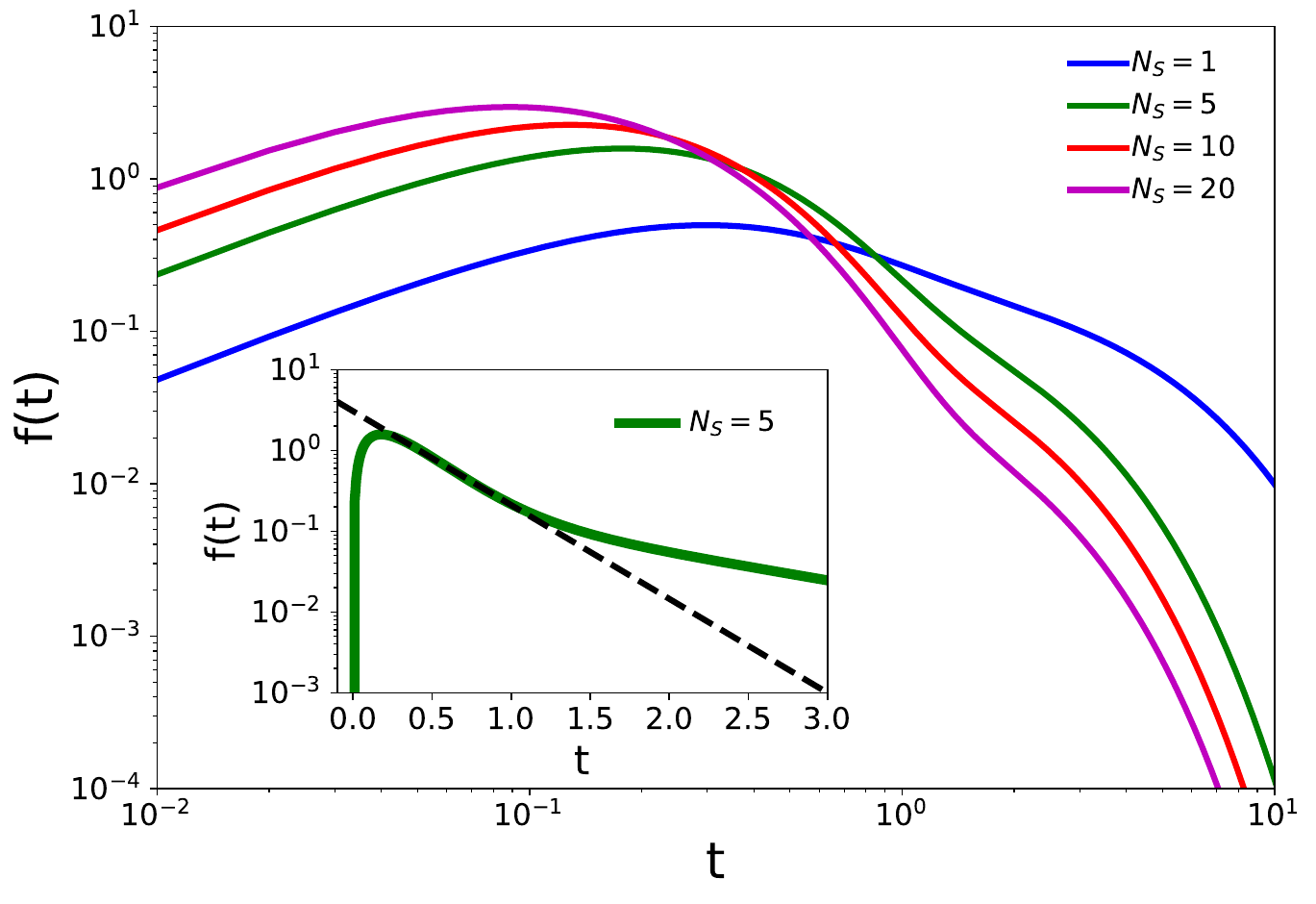}} &
    \subcaptionbox{Uncompetitive\label{fig:FPT-uncompetitive}}[0.47\linewidth]{\includegraphics[width=\linewidth]{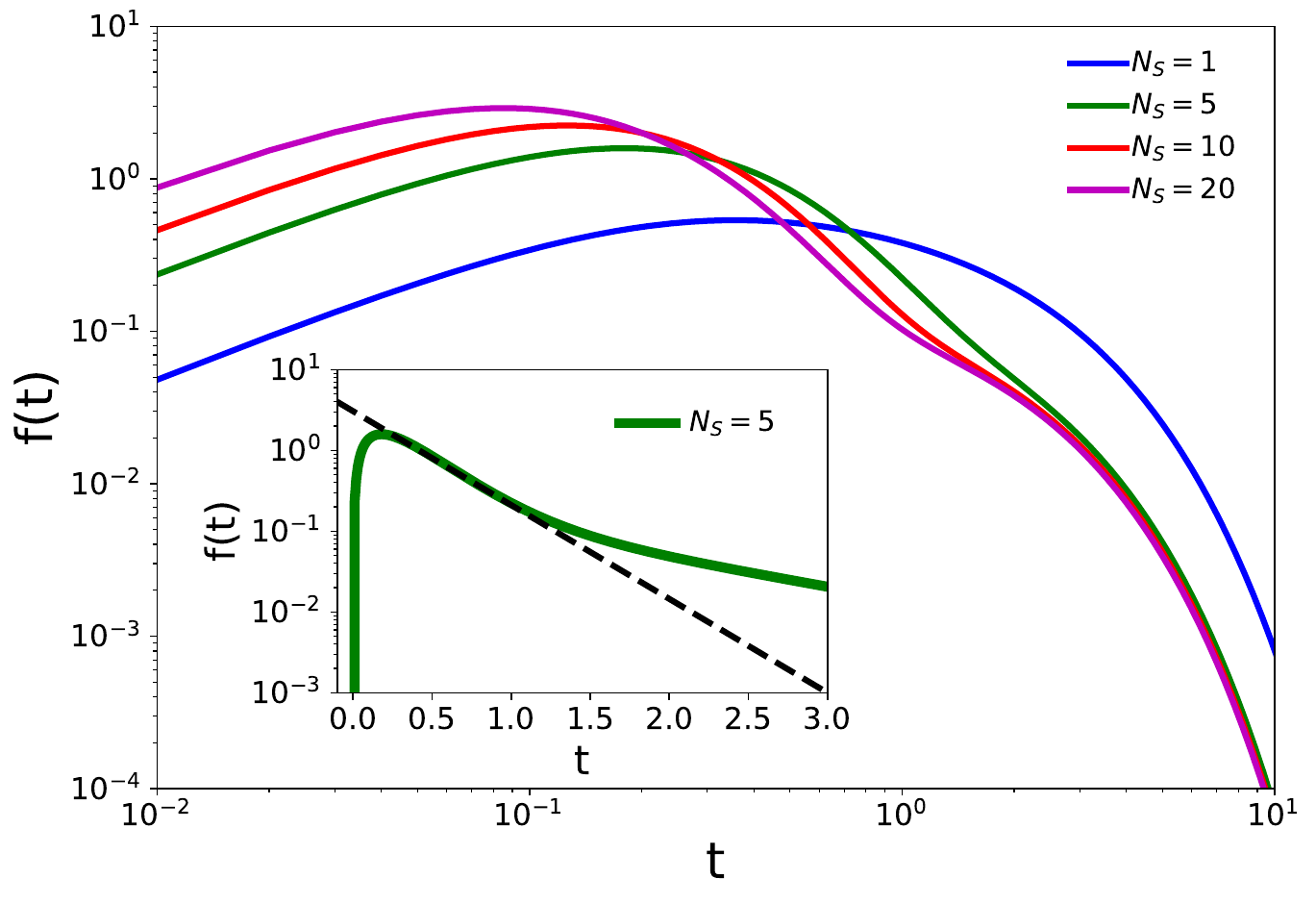}} \\
    \subcaptionbox{Noncompetitive\label{fig:FPT-noncompetitive}}[0.47\linewidth]{\includegraphics[width=\linewidth]{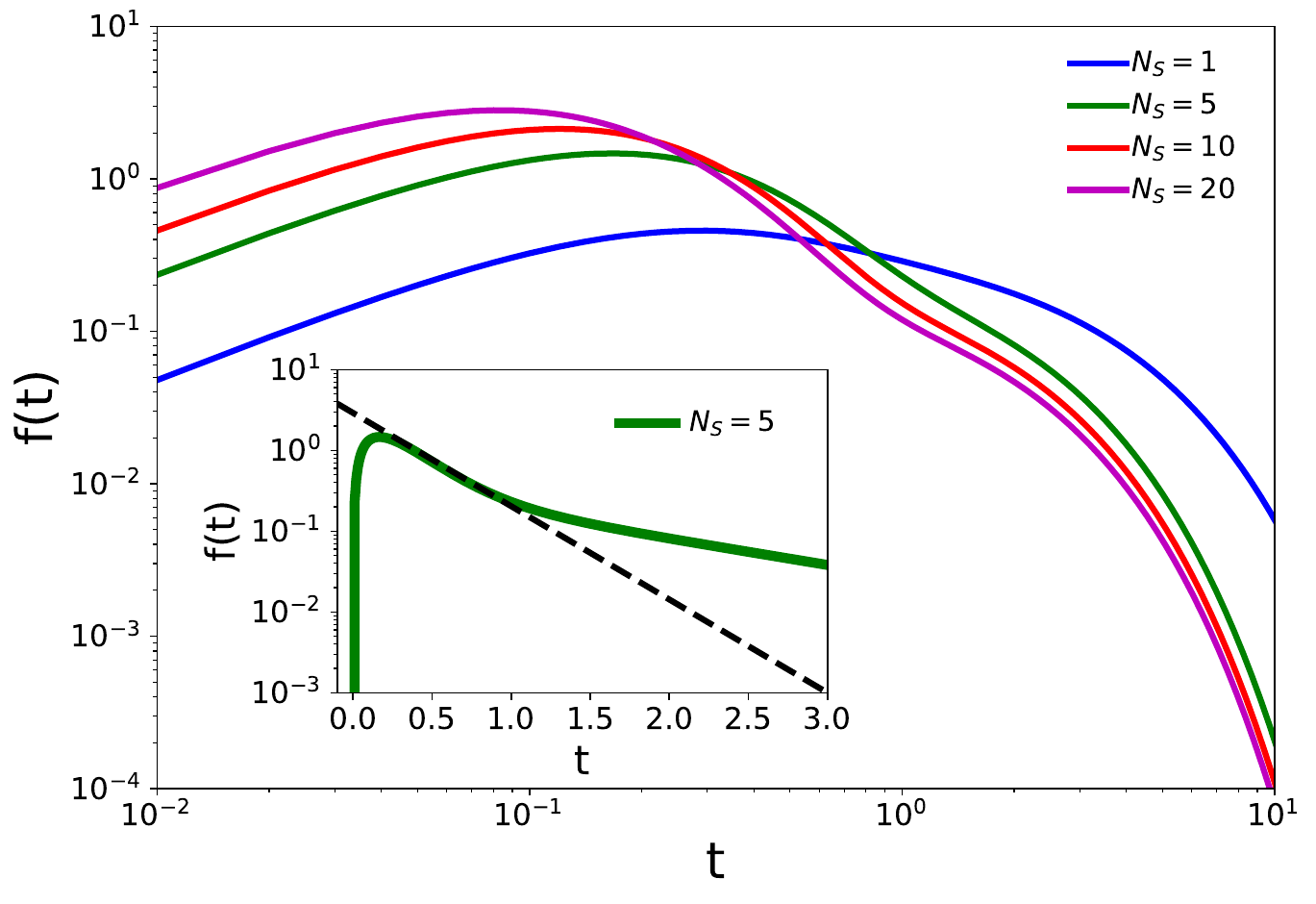}} &
    \subcaptionbox{Partial\label{fig:FPT-partial}}[0.47\linewidth]{\includegraphics[width=\linewidth]{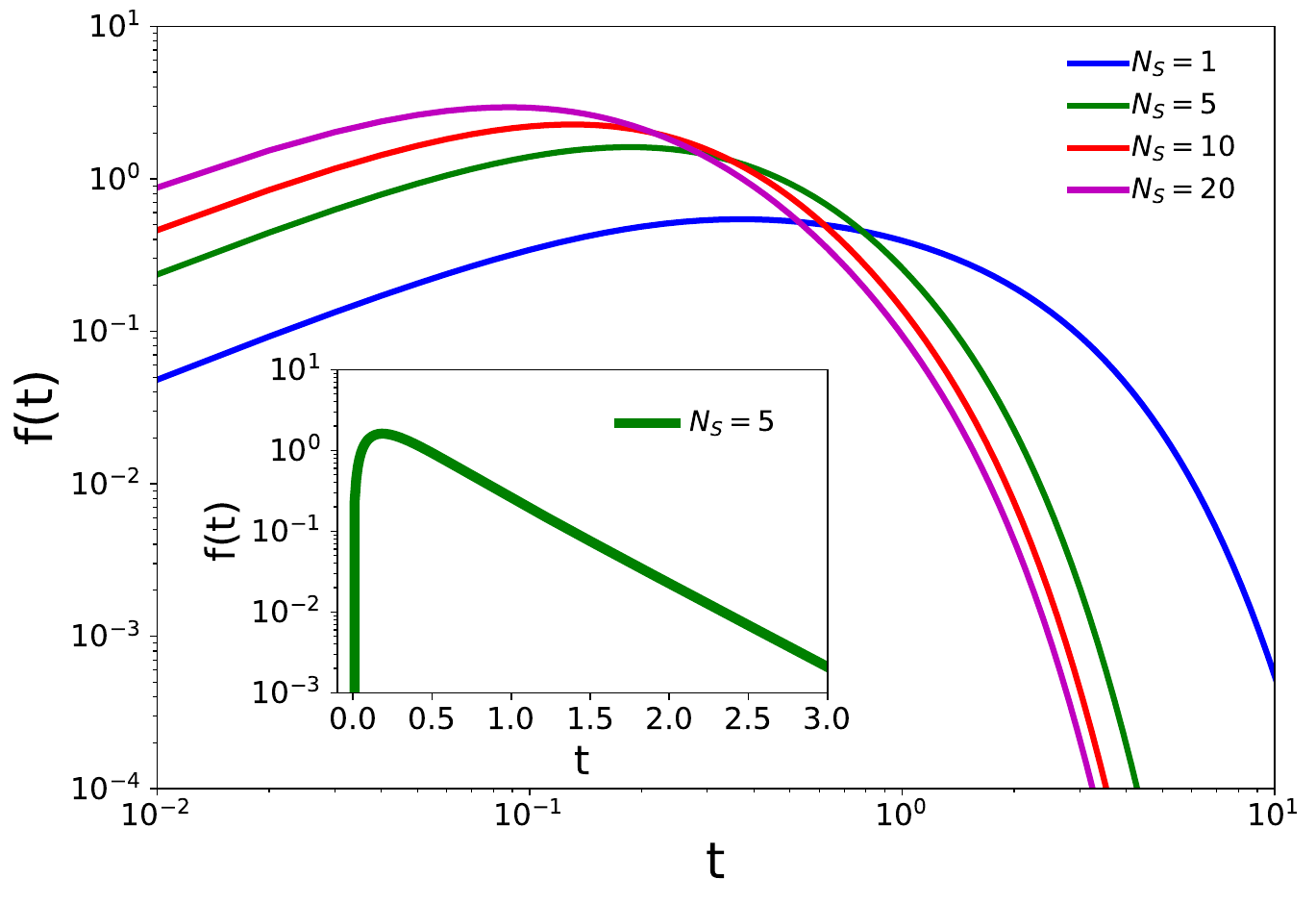}}
  \end{tabular}
  \caption{The probability density of the first formation of the product as a function of $t$ on a log-log scale, for different values of $N_S$, $N_E = N_I = 1$ and rate constants: (a) $k_{1\pm}=k_{3\pm}=1$ and $k_2=5$, (b) $k_{1\pm}=k_{4\pm}=1$ and $k_2=5$, (c) $k_{1\pm}=k_{3\pm}=k_{4\pm}=k_{5\pm}=1$ and $k_2=5$, (d) $k_{1\pm}=k_{3\pm}=k_{4\pm}=k_{5\pm}=1$, $k_2=5$ and $k_6=2.5$.
   Inset: $f(t)$ on a log-normal scale. The dashed line serves as a guide-to-the-eye highlighting the intermediate time scale.}
  \label{fig:FPT_inhibition}
\end{figure*}


Figure \ref{fig:FPT_inhibition} presents the results for competitive, uncompetitive, noncompetitive and partial inhibitions for $N_E=N_I=1$ and $N_S=\{1,5,10,20\}$. 
In the inset, we present the results with $N_S=5$ on log-normal scale. 
Our results suggest the existence of an intermediate time scale for the competitive, uncompetitive and noncompetitive inhibitions highlighted by dashed lines in the insets.


The FPFT distribution can be written as follows
\begin{equation}
    f(t) = \sum_{i=1}^{N^{\prime}} \alpha_i e^{-\bar{\lambda}_i t}, 
    \label{eq:series_FPT}
\end{equation}
where the set $\{\bar{\lambda}_i| i=1\ldots N^{\prime}\}$ represents the nonzero eigenvalues organized in ascending order
and $\sum^{N^{\prime}}_{i=1} \alpha_i=0$. For the simplest case $N_S=N_E=N_I=1$ discussed above, we have $\alpha_1=- \alpha_3=kk^{\prime}/\sqrt{\Delta}$ and $\alpha_2=0$. In the short-time regime $t\approx 0$, we can expand Eq. (\ref{eq:series_FPT}) in a power series given by $f(t\approx 0) = -\sum_{i=1}^{N^{\prime}} \alpha_i \bar{\lambda}_i t + \mathcal{O}(t^2)$. This linear behavior can be seen in the main plots of Fig. \ref{fig:FPT_inhibition} for $t\approx 0$. 

Expanding around an intermediate time $t=t_0$, we find
\begin{equation}
    f(t) = A_0\left[1-B_0(t-t_0)\right] + \mathcal{O}\left((t-t_0)^2\right),
    \label{eq:FPT_expansion}
\end{equation}
where $A_0 = \sum^{N^{\prime}}_{i=1} \alpha_ie^{-\bar{\lambda}_it_0}$ and $B_0=\left(\sum^{N^{\prime}}_{i=1} \alpha_i\bar{\lambda}_ie^{-\bar{\lambda}_it_0}\right)/A_0$. Taking the logarithm of $f(t)$ we get 
\begin{equation}
    \ln f(t) \approx \ln A_0 - B_0(t-t_0).
    \label{eq:FPT_log_expansion}
\end{equation}
For $t > 2$ in panels (a) to (c) of Fig. \ref{fig:FPT_inhibition}, only the term related to the lowest eigenvalue is relevant, thus Eq. (\ref{eq:FPT_log_expansion}) reduces to 
$\ln f(t>2) \approx \ln \alpha_1 - \bar{\lambda}_1t$ 
as expected. Therefore, as discussed in the simplest case, the lowest eigenvalue sets the long-term time scale.

The results presented in panels (a) to (c) reveal the appearance of an intermediate time scale related to the slope $B_0$ for $0.5 \lesssim t \lesssim 1.5$. 
This intermediate time scale is characterized by the contribution of other eigenvalues instead of only $\bar{\lambda}_1$. 
For the partial inhibition kinetic reaction, shown in panel (d), $B_0 = \bar{\lambda}_1$ for $t \approx 0.5$, and therefore there are only two time scales in this case, similar to the regular Michaelis-Menten kinetic reaction. 

The intermediate time scale also depends on the number of substances that participate in the reaction. As we can see in the main plots of panels (a) to (c), 
the intermediate time scale becomes more evident as $N_S$ increases. We also note that the slope is approximately the same for the inhibitions considered $B_0\approx 1$, see the insets of panels (a) to (c). 
This suggests that the delay in the formation of the product caused by the interaction of the reactants with the inhibitor is responsible for the appearance of this intermediate time scale,  
independently of the specific inhibitory mechanism considered. 

\begin{figure*}
  \begin{tabular}{ccc}
    \subcaptionbox{Competitive \label{fig:FPT-competitive2}}[0.33\linewidth]{\includegraphics[width=\linewidth]{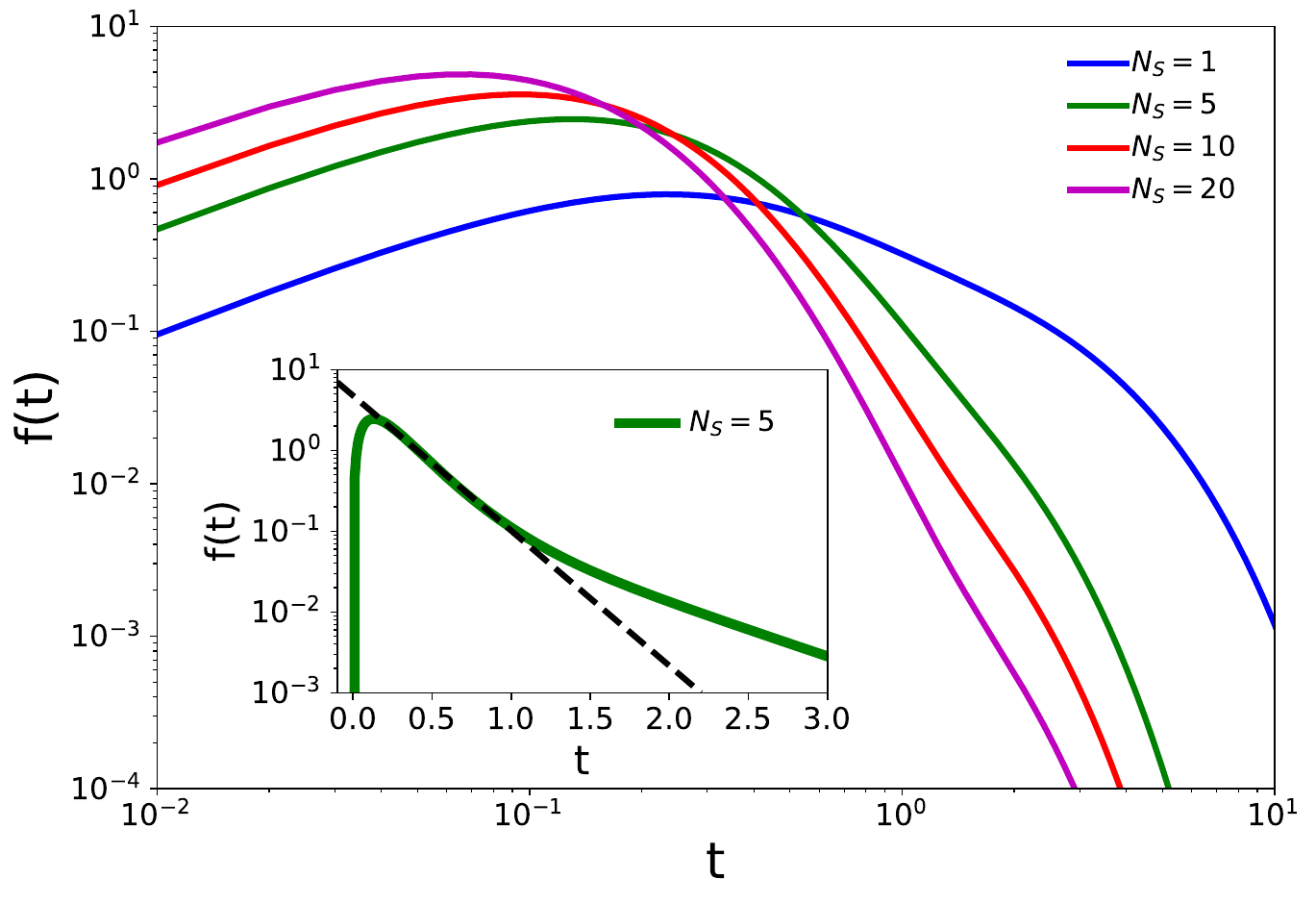}} &
    \subcaptionbox{Uncompetitive\label{fig:FPT-uncompetitive2}}[0.33\linewidth]{\includegraphics[width=\linewidth]{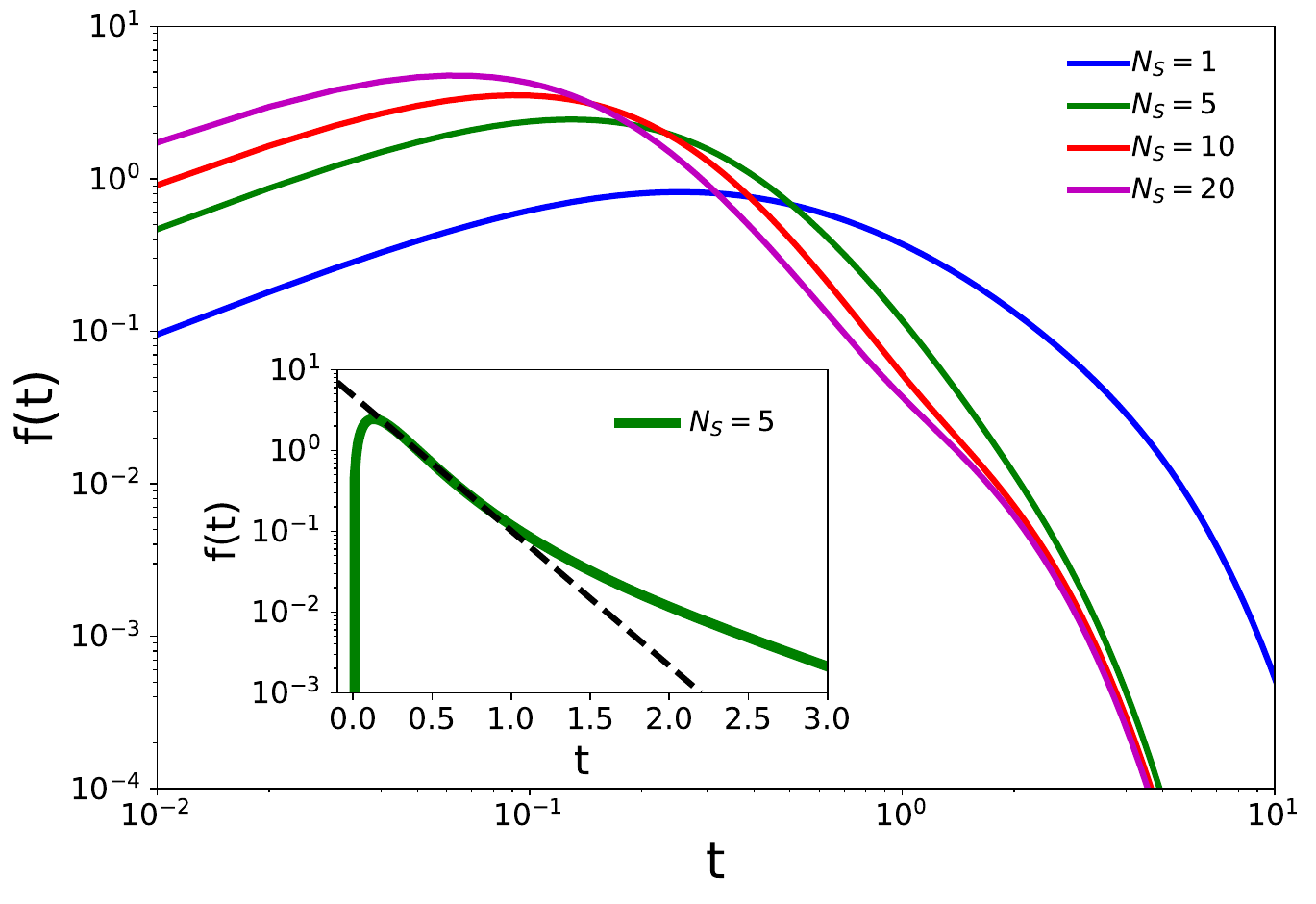}} &
    \subcaptionbox{Noncompetitive \label{fig:FPT-noncompetitive2}}[0.33\linewidth]{\includegraphics[width=\linewidth]{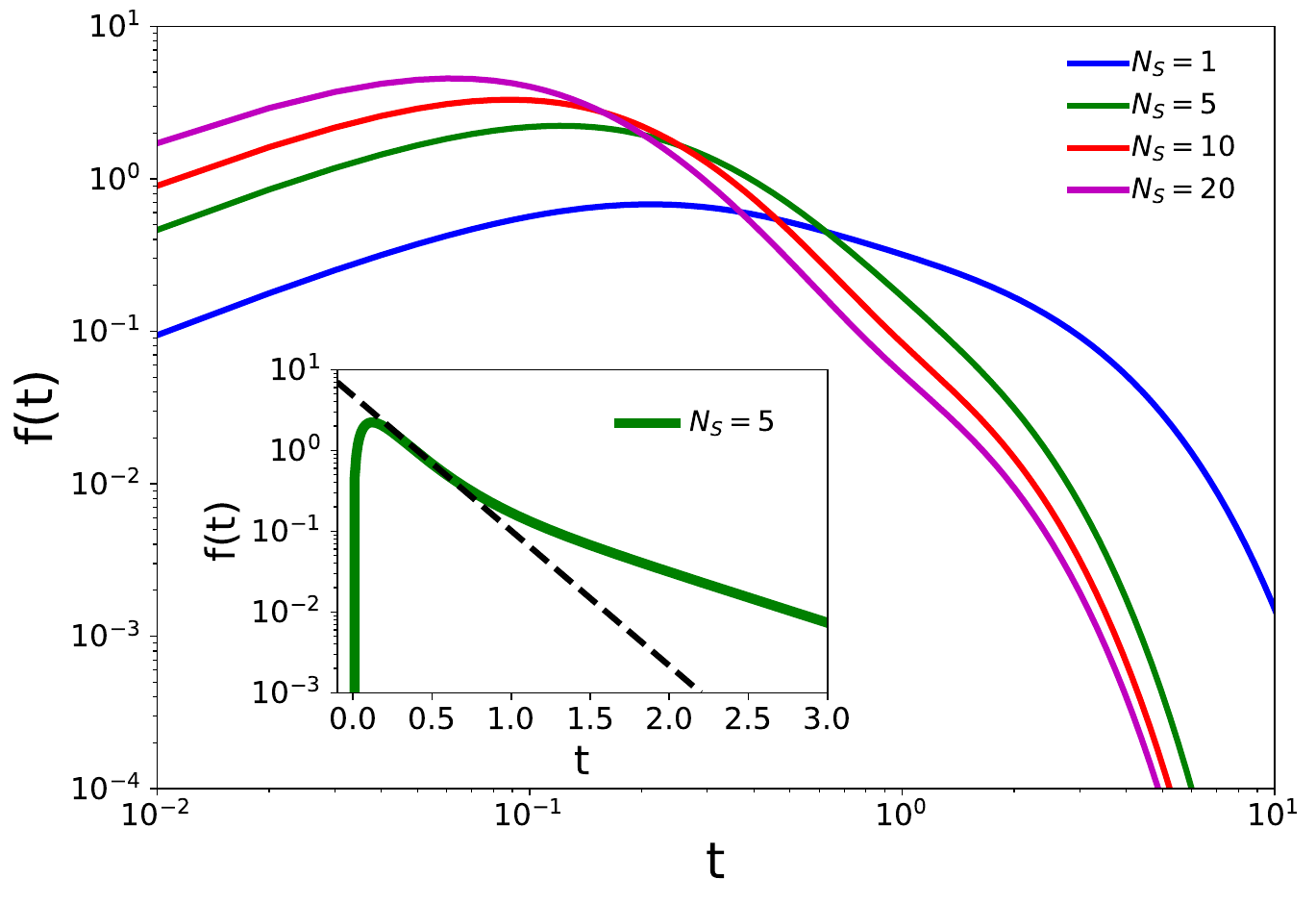}}
  \end{tabular}
  \caption{The probability density of the first formation of the product as a function of $t$ on a log-log scale, for different values of $N_S$, $N_E = N_I = 2$ and rate constants: (a) $k_{1\pm}=k_{3\pm}=1$ and $k_2=5$, (b) $k_{1\pm}=k_{4\pm}=1$ and $k_2=5$, (c) $k_{1\pm}=k_{3\pm}=k_{4\pm}=k_{5\pm}=1$ and $k_2=5$. Inset: $f(t)$ on a log-normal scale. The dashed line serves as a guide-to-the-eye highlighting the intermediate time scale.}
  \label{fig:FPT_inhibition_NE=NI=2}
\end{figure*}

Figure \ref{fig:FPT_inhibition_NE=NI=2} illustrates the FPFT distribution for the competitive, uncompetitive, and noncompetitive inhibitions with both $N_E$ and $N_I$ set to 2, while varying $N_S$. This figure demonstrates that the intermediate scale is maintained even when the number of enzymes is increased. Furthermore, it is important to note that, in this scenario, the slope of the dashed line is steeper than that seen in the single-enzyme case. In this case, $B_0$ is roughly three for all inhibitory mechanisms considered. 

In Ref. \cite{godec2016first} the authors examine the first-passage time of 
Brownian motion in a confined spherically symmetric reflecting boundary with a heterogeneous diffusion coefficient and a central target. 
The paper particularly highlights the emergence of a previously overlooked third time scale. This new time scale is characterized by brief excursions away from the target before the particle returns, which adds complexity to the understanding of particle trajectories. 
Parallels can be drawn between these brief excursions with intermediate reactions caused by the interaction of the enzyme and the complex $C_1$ with inhibitors that delay the formation of the first product in our kinetic reaction. 
Our findings emphasize the importance of examining the entire distribution rather than solely focusing on its moments. Distributions tend to be more informative than average quantities, such as the mean formation time of the first product, which may not provide an accurate representation of the behavior of the system \cite{Mejia_Monasterio:JSM2011,Mattos_etal:pre2012}.

Although intermediate time scales have been empirically reported in a variety of systems \cite{thorneywork2020direct,saha2012single}, and predicted for branched kinetic networks \cite{li2013mechanisms,valleriani-etal:jcp2014}, our analysis provides the first closed-form derivation of a textbook enzymatic mechanism.  The simplicity of the inhibited Michaelis–Menten scheme therefore makes it an attractive test-bed for the general ideas of Refs. \cite{godec2016first,li2013mechanisms,valleriani-etal:jcp2014}.

\section{\label{sec:conclusion} Conclusion}

In this work, we analyze the stochastic Michaelis-Menten kinetic reaction with four different types of inhibition mechanism namely, competitive, noncompetitive, uncompetitive, and partial inhibitions.
We map the master equation of this stochastic process as a Schrödinger equation using the Fock space approach. 
The probability evolution of a specific configuration of the system occurring in a given time $t$ is dictated by the quasi-Hamiltonian operator Eq. (\ref{eq:quasi-hamiltonian}), written in terms of creation and annihilation ladder operators.

We obtain simple analytical expressions for the probabilities associated with each possible configuration of the simplest case. In this scenario, only an enzyme, a substrate, and an inhibitor ($N_S=N_E=N_I=1$) participate in the reaction. This demonstrates the power of the Fock space formalism in studying stochastic small copies systems.
Similarly to the usual Michaelis-Menten kinetic reaction, the stiffness is also present when inhibitory mechanisms are taken into account. 

Stiffness is a characteristic of systems with more than one timescale. It is well known that numerical simulations can be very demanding in systems with multi-timescales. Our method handles this issue well even with systems containing around 12 substances. 
In the case where we have $N_S=6$ and $N_E=N_I=3$, we analyze the overall role of inhibitors in the formation of products in partial inhibition.
We observed that inhibitors do not affect product formation when $k_2=k_6$,
and for $k_2<k_6$ the inhibitor works as an activator, facilitating product formation rather than hindering it. 

Additionally, we explore the problem of estimating the time necessary for the formation of the first product in Michaelis-Menten kinetics with inhibition, interpreted as a first-passage process.
Typically, the first-passage time distribution is characterized by a two-time-scale behavior such as the usual Michaelis-Menten kinetics.
We obtain an expression for the first product formation time (FPFT) distribution Eq. (\ref{eq:FPT_multiple}) for Michaelis-Menten with inhibition using the Fock space approach.
We used the Lineweaver-Burk linearization to compare our result with those found in the literature, which shows excellent agreement. 

We observe the emergence of an intermediate time scale for the competitive, uncompetitive, and noncompetitive inhibition mechanisms. This new time scale seems to be approximately the same for the inhibitions considered and becomes more evident as the number of substances increases.
This intermediate time scale refers to the enzyme and complex $C_1$ interactions with the inhibitor that slow the formation of the first product in our kinetic reaction.

This result indicates how crucial knowledge of distributions such as FPFT is in various fields, such as chemical reactions, intracellular transport, 
materials science, etc.
Distributions often convey more information than moments, such as the mean formation time of the first product, which may not accurately reflect the system behavior.
By recognizing intermediate time scales, researchers can gain deeper insight into the kinetics of first-passage processes, leading to more accurate models and predictions in complex systems.

Building upon this work, we aim to elucidate the underlying mechanisms responsible for the appearance of additional time scales in stochastic processes such as the kinetic reactions studied here. Recently, the H theory was introduced as a framework for understanding non-Gaussian behavior in hierarchical stochastic processes \cite{salazar-vasconcelos:pre2012,macedoetal:pre2017}. While this theory has so far been applied primarily to describe the asymptotic behavior of heavy-tailed fluctuations in various systems \cite{gonzalez-etal:naturecomm2017,sosa-correa-etal:prfluids2019,barbosa-etal:prl2022,vasconcelos-etal:pre2024}, we propose that this framework can elucidate the mechanisms responsible for the appearance of intermediate time scales and reveal possible interplays between them.

\begin{acknowledgments}
This work was supported by Coordenação de Aperfeiçoamento de Pessoal de Nível Superior (CAPES).
The authors thank Prof. Anderson L. R. Barbosa for fruitful discussions regarding the present work.
\end{acknowledgments}




\bibliographystyle{plain}
\bibliography{apssamp}

\end{document}